\documentclass[a4paper,conference]{IEEEtran}
\IEEEoverridecommandlockouts
\usepackage{subcaption}
\usepackage{cite}
\usepackage{amsmath,amssymb,amsfonts}
\usepackage{algorithmic}
\usepackage{graphicx}
\usepackage{textcomp}
\usepackage{balance}
\usepackage{graphicx,multirow}
\usepackage[table]{xcolor}
\usepackage[nomessages]{fp}
\usepackage{graphicx,multirow}
\usepackage{array}
\usepackage[table]{xcolor}
\usepackage{array}
\usepackage{pgfplots, pgfplotstable}
\usetikzlibrary{patterns}
\usepackage{multirow}
\usepackage{adjustbox}
\def\BibTeX{{\rm B\kern-.05em{\sc i\kern-.025em b}\kern-.08em
		T\kern-.1667em\lower.7ex\hbox{E}\kern-.125emX}}
\begin{document}
	
	\title{Non-Intrusive Load Monitoring (NILM) using Deep Neural Networks: A Review}
	
	\author{\IEEEauthorblockN{Mohammad Irani Azad}
		\IEEEauthorblockA{\textit{Faculty of ECE} \\
			\textit{Qom University of Technology}\\
			Qom, Iran \\
			iraniazad.m@qut.ac.ir}
		\and
		\IEEEauthorblockN{Roozbeh Rajabi}
		\IEEEauthorblockA{\textit{Faculty of ECE} \\
			\textit{Qom University of Technology}\\
			Qom, Iran \\
			rajabi@qut.ac.ir}
		\and
		\IEEEauthorblockN{Abouzar Estebsari}
		\IEEEauthorblockA{\textit{School of the Built Environment and Architecture} \\
			\textit{London South Bank University}\\
			London, United Kingdom \\
			estebsaa@lsbu.ac.uk}
	}
	
	\IEEEoverridecommandlockouts
	
	\maketitle
	
	\IEEEpubidadjcol

	\begin{abstract}
		Demand-side management now encompasses more residential loads. To efficiently apply demand response strategies, it's essential to periodically observe the contribution of various domestic appliances to total energy consumption. Non-intrusive load monitoring (NILM), also known as load disaggregation, is a method for decomposing the total energy consumption profile into individual appliance load profiles within the household. It has multiple applications in demand-side management, energy consumption monitoring, and analysis. Various methods, including machine learning and deep learning, have been used to implement and improve NILM algorithms. This paper reviews some recent NILM methods based on deep learning and introduces the most accurate methods for residential loads. It summarizes public databases for NILM evaluation and compares methods using standard performance metrics.
	\end{abstract}
	
	
	\section{Introduction}
	
	The non-intrusive load monitoring (NILM) method has gained popularity in recent years as a way to monitor appliance and electrical utility energy usage in buildings and events (on/off) using a single energy meter. If consumers had data on appliance-level energy usage, they could better understand their energy consumption behavior and take action to reduce it. The aim of this study is to present an overview of the latest algorithms currently being investigated by researchers to create a precise non-intrusive load monitoring (NILM) method for effective energy management. The article discusses the potential applications of NILM across different fields, along with future research objectives. The development of sustainable and smart cities has been made possible by advancements in artificial intelligence (AI), smart meters, the internet of things (IoT), and smart grids, as cited in \cite{RN1} and \cite{RN2}. Effective energy management is a crucial component of sustainable city development, which aims to utilize resources responsibly, protect the environment, and enhance society's well-being. The objective of energy management is to promote energy system self-reliance and sustainability \cite{RN1}.
	
	Energy management involves monitoring and controlling electrical utilities to optimize energy use and reduce consumption. However, with the increase in energy needs, energy conservation has become a challenge in recent years \cite{RN3}. Greater energy use can lead to an energy crisis, climate change, and a negative impact on the economy \cite{EEEIC_PriceGRU}. It is estimated that the rise in carbon emissions will increase global temperatures by 2.5 to 10 $^{\circ}$C this century, causing more frequent floods, droughts, a rise in sea level, and the spread of infectious illnesses \cite{RN4}. Therefore, it is essential to reduce carbon emissions across all sectors, including construction, industry, and transportation, to mitigate climate change. Researchers are working on developing technology solutions for energy conservation \cite{RN3}. Buildings are one of the major contributors to energy consumption \cite{RN5}, with energy consumption in this sector steadily increasing over time. In order to mitigate carbon emissions, optimizing energy consumption in residential and commercial buildings is crucial. This can be achieved through the construction or design of energy-efficient structures, as well as improving energy usage in existing buildings.
	
	The paper is organized as follows. Section \ref{sec:review} introduces the mathematical definition of the NILM problem. Section \ref{sec:deepMethods} discusses deep learning-based NILM methods. Section \ref{sec:datasets} provides a summary of the public NILM datasets. Section \ref{sec:experiments} presents a comparison study of NILM methods, and finally, Section \ref{sec:conclusion} concludes the paper.
	
	\section{NILM Problem Definition}\label{sec:review}
	\subsection{Mathematical Problem Definition}\label{sec:math}
	
	The issue at hand can be described as follows: at a given time $t$, the total active power consumed by a system is represented by $y(t)$, while $y_i(t)$ represents the active power consumed by the $i$th appliance at the same time. The overall load is the sum of the energy consumed by individual appliances and an unmeasured residual load, expressed as:
	
	\begin{equation}
		y(t) = \sum_{i=1}^{N} y_i(t) + e(t),
	\end{equation}
	
	where $N$ denotes the number of appliances considered, and $e(t)$ represents the undetermined residual load. The aim is to estimate $F(y(t))$ by determining the values of $y_i(t)$, given only the value of $y(t)$, as:
	
	\begin{equation}
		y_1(t), y_2(t), ..., y_i(t), ..., y_N(t) = F(y(t)),
	\end{equation}
	
	where $F$ is an operator that produces $N$ distinct values when applied to the total active power. These numbers represent the most accurate estimate of the power consumed by each appliance. It should be noted that $y_i(t)$ typically does not reflect the entire set of home appliances but rather a subset of them. As a result, the unknown term $e(t)$ takes into account the loads caused by unmonitored appliances. If simultaneous measurements of the aggregate consumption and load of each appliance are available, approximating the $F$ operator can be considered a supervised learning problem. When mainly concerned with activation times and cumulative consumption, as is the case in real situations, the estimated individual appliance consumption ($\hat{y}_i(t)$) can be obtained using functions that are constant over the device's activation period:
	
	\begin{equation}
		\hat{y}_i(t) = p_i\hat{a}_i(t),
		\label{eq:constants}
	\end{equation}
	
	where $p_i$ represents the average consumption of appliance $i$, and $\hat{a}_i(t)$ represents an estimate of the activation state of the particular appliance at time $t$. Its value is one if the device is in use and uses energy, and zero otherwise. Therefore, starting with the aggregate load, a technique is provided to derive the most accurate and feasible assessment of the activation state of the appliances:
	
	\begin{equation}
		\hat{a}_1(t),\hat{a}_2(t),...,\hat{a}_i(t),...,\hat{a}_N(t) = F_a(y(t)),
	\end{equation}
	
	After learning the average nominal consumption of the considered equipment, one can use Equation \ref{eq:constants} to estimate consumption.
	
	\subsection{Appliance types}\label{sec:types}
\begin{figure}[tb]
		\centerline{\includegraphics[width=8cm]{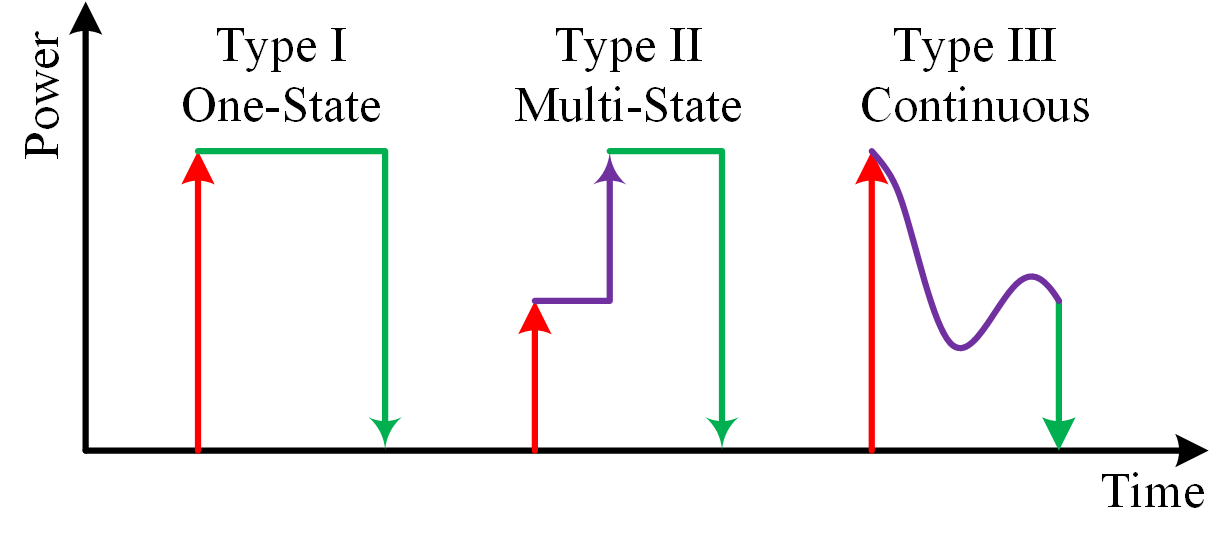}}
		\caption{A pictorial representation illustrates the different types of appliances categorized based on their operating states.}
		\label{fig:ApplianceTypes}
	\end{figure}

	Based on their operational characteristics, appliances can be classified into four types as discussed in \cite{RN12}. Type I appliances have two modes of operation - on and off. These include appliances such as kettles, toasters, and light bulbs, which consume energy only when turned on. Type I appliances are predominantly resistive with few linear reactive components. Type II appliances are characterized as multi-state or finite state machines with a limited number of operational states that may be run repeatedly. Changes in these appliances' states can be observed by monitoring the power consumption's falling/rising edges over time. Stove burners, refrigerators, and washing machines are some examples of Type II appliances \cite{RN12,RN13}. Figure \ref{fig:ApplianceTypes} demonstrates the distinct appliance operation conditions.
	
	Category III appliances, also known as Continuously Variable Devices (CVDs), exhibit a non-repetitive power usage pattern, which poses a challenge for energy consumption disaggregation. Examples of Type III appliances include power drills and dimmer lights \cite{RN13}. Type IV appliances are those that run continuously for extended periods of time, typically lasting several days or weeks. Examples of Type IV equipment include wireless telephone devices and cable TV receivers \cite{RN13}. Therefore, the Non-Intrusive Load Monitoring (NILM) system is required to have the ability to differentiate between various types of appliance events, which may happen concurrently or independently and at varying time intervals.
	
	\section{Deep Learning Based NILM methods}\label{sec:deepMethods}
	NILM techniques can be broadly categorized into two groups: supervised and unsupervised methods \cite{RN18}. In supervised NILM, individual appliance power usage is used to train the models. On the other hand, unsupervised methods can only utilize aggregate power usage data. Examples of unsupervised NILM techniques include Hidden Markov models (HMM) \cite{RN19,RN20}, factorial HMM (FHMM) \cite{RN21,RN14}, and techniques based on event detection and clustering \cite{RN22,RN15}. These techniques have been thoroughly examined in previous studies \cite{RN18,RN23}. With the advent of deep neural networks (DNNs), many neural network-based supervised NILM techniques have been developed \cite{RN24,RN25}. Convolutional neural networks (CNN) have also recently made significant advances \cite{RN26,RN27}. Graph signal processing \cite{RN10}, HMM \cite{RN21,RN29,RN14,RN15,RN30}, and DNNs \cite{RN31,RN32} are commonly used in suggested NILM approaches. As the cost of employing appliance data for training has grown dramatically, researchers have focused on developing unsupervised approaches and incorporating appliance models. Despite the significant progress made in NILM research in recent years, challenges remain in terms of application, identification accuracy, training time, and online deployment techniques in smart metering frameworks.
	
	\subsection{Event-Based Non-Intrusive Detection}
\begin{figure}[tb]
		\centerline{\includegraphics[width=2.5cm]{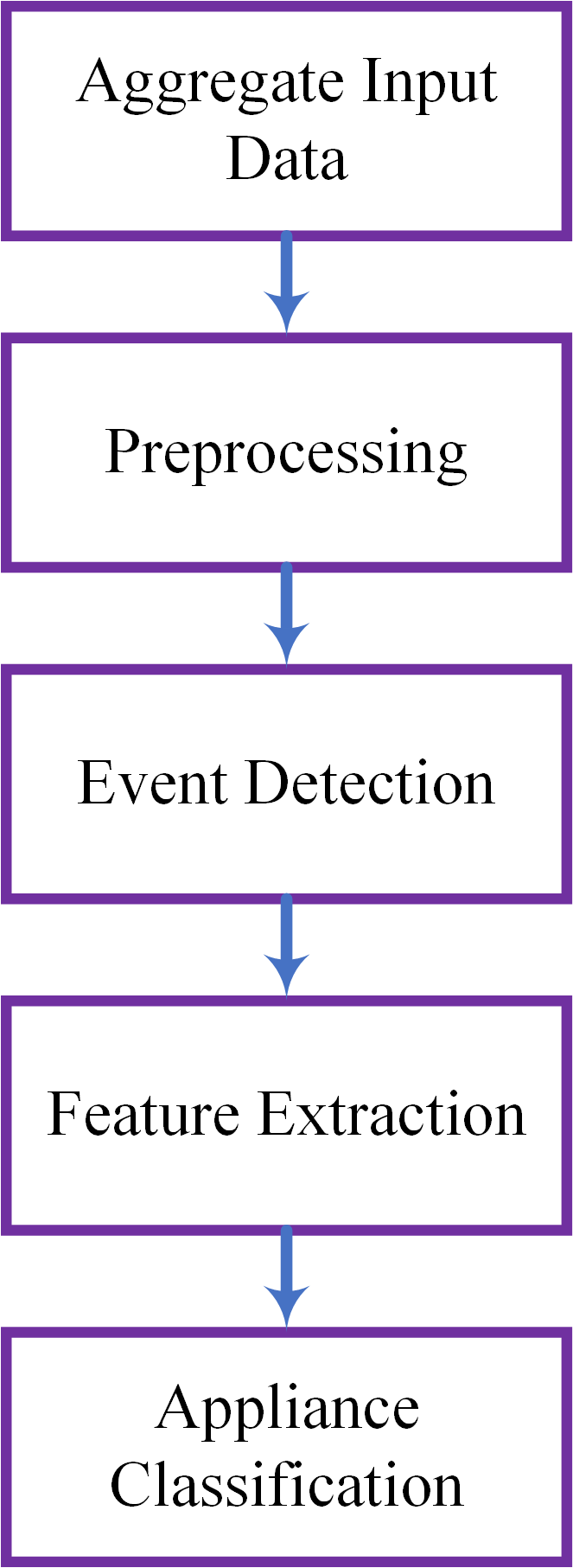}}
		\caption{Block diagram scheme of event-based NILM.}
		\label{fig:event-based}
	\end{figure}

	The event-based NILM method is based on the concept of detecting and categorizing events within a combined electrical signal. Figure \ref{fig:event-based} shows the block diagram of the this approach. A robust event detector should be developed to cope with noisy fluctuations and identify events with decay and growth patterns, which is a bottleneck and inherent difficulty in existing event detectors \cite{RN33}. One approach includes the steps of event detection, extraction, clustering, and matching in the event-based block \cite{onlineNILM}. It should be noted that the accuracy of previous event-based frameworks is dependent on the power features that are introduced. Since some appliances may have identical active power curves but radically distinct reactive power trends, increasing the number of features can enhance the accuracy of the appliance model, particularly for non-linear loads. One of the advantages of incorporating reactive power is that it enables discrimination between different types of devices.
	
	\subsection{NILM Disaggregation by CNN}
	
	The proposed method \cite{MDPI_CNN} employs a convolutional neural network (CNN) that takes the time interval of a home's energy consumption as input and predicts the activation status of each device at every time step. The network architecture, referred to as Temporal Pooling NILM (TP-NILM), is an updated version of Zhao et al.'s PSPNet (Pyramid Scene Parsing Network) used for semantic image segmentation \cite{RN34}. The TP-NILM follows the conventional approach to image segmentation, with an encoder comprising pooling and convolutional layers that enhance the feature space of the signal but decrease its temporal resolution, and a decoder module that uses these features to approximate the activation state of the devices at the original resolution. To establish a temporal context that covers extended periods without compromising the signal's resolution, the TP-NILM incorporates a Temporal Pooling module that accumulates features at various resolutions, enabling accurate reconstruction of the activation state.
	
	\begin{figure}[htbp]
		\centerline{\includegraphics[width=4.25cm]{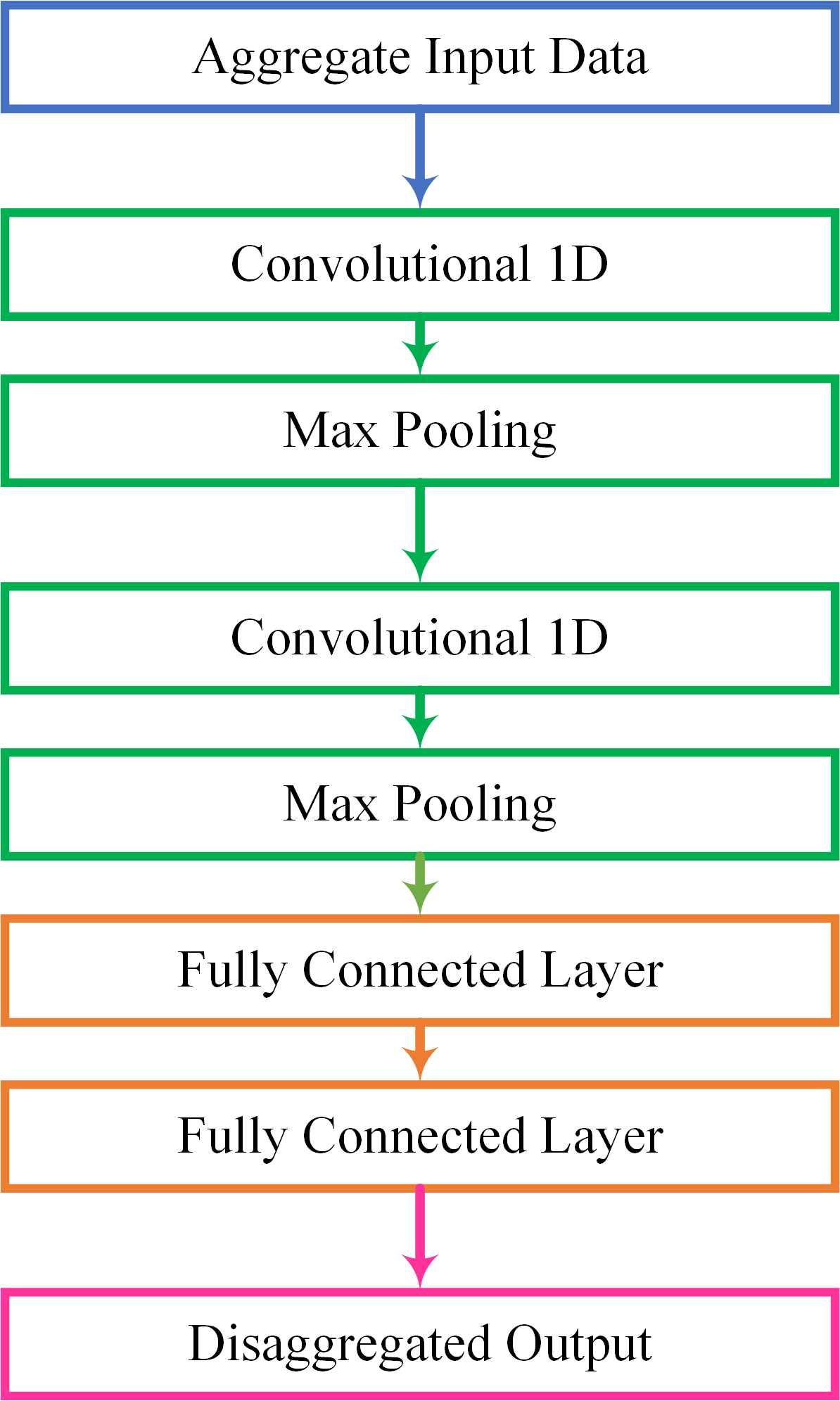}}
		\caption{Outline of network structure for NILM by CNN.}
		\label{fig:NILMbyCNN}
	\end{figure}

	Figure \ref{fig:NILMbyCNN} illustrates the architecture of the network used in this study. The encoder uses a rectified linear unit activation function, batch-independent normalization downstream of the activations, and a regularization dropout layer, with three convolutional filters interleaved by max-pooling layers. The encoder reduces the signal's temporal resolution by a factor of eight and raises the number of output characteristics from a single aggregate power consumption value to 256. The TP block provides context information to the decoding block, allowing it to create extra features for decoding by aggregating encoder output with various resolutions. The encoder output is passed through four average pooling modules with various filter sizes, which degrade the temporal resolution while maintaining the number of features, before being convolutioned with a unit filter size. This reduces the number of characteristics to one-quarter of those used in the input. The convolutional results are used as the input for a rectified linear unit activation function, which is then batch normalized. Lastly, linear up-sampling yields a temporal resolution at the TP block's output that is equal to the encoder's output. A dropout is also added to the output, allowing the network to be controlled. The context features generated by this block are connected to the encoder's detail features, doubling the total number of features in the decoder's input. To raise the signal's temporal resolution and lower the number of features, the decoder contains a transposed convolutional layer with a stride and kernel size of 8. The ReLU is still used as the activation function, followed by an extra convolutional layer with an unified kernel size that keeps the temporal resolution and increases the number of output channels to match the number of devices being analyzed. In the output, a sigmoid function is used. This is because in the semantic segmentation of photos, each pixel is connected to a single class, however in the current application, many appliances might be in use at the same time. While working in this manner, the network decomposes all appliances at the same time. This should enable an encoder to use more broad convolutive filters that aren't specialized for a single kind of appliance, boosting the neural network's capacity to generalize. Gradient descent optimization may be used to find the net's weights. The loss function is a binary crossentropy applied to each output channel that assesses the disparity between the activations predicted by the net $ \hat{a}_i(t) $ and the actual ones $ {a}_i(t) $ for each appliance under consideration and for each instant of the period under consideration.

	\section{NILM Public Datasets}\label{sec:datasets}
	To develop NILM (non-intrusive load monitoring) algorithms and assess their performance, the research community provides various NILM datasets in the public domain \cite{RN35}. Since each dataset monitors different appliances in diverse environments and buildings over a varied time period, each dataset has its own specific criteria \cite{RN36,RN37}. However, it has been observed that many public datasets have structural variations that require pre-processing before usage. To address this issue, the dsCleaner Python module was developed to standardize, clean, and convert time series data into a consistent file format, and also includes a resampling method for datasets. Typically, NILM datasets consist of aggregated energy data from a single meter and the actual energy consumption of each appliance, which is measured by plug-level meters and serves as the ground truth for evaluating NILM algorithms. Table \ref{tbl:comparison} lists the most popular publicly accessible NILM datasets for research purposes \cite{RN35}.

	\begin{table*}[tb]
		\caption{Summary of NILM public datasets}
		\begin{center}
			\begin{tabular}{|c|p{1.5cm}|c|c|p{1.5cm}|p{1.5cm}|c|p{5.5cm}|}
				\hline
				\textbf{Dataset}&{Sampling Rate Time Resolution}&{Duration}&{Type}&{Number of Households}&{Number of Appliances}&{Country}& {URL} \\
				\hline
				\textbf{REDD} & 16.5 KHz & 19 days & Residential & 6 & 6 & US & {https://tokhub.github.io/dbecd/links/redd.html} \\
				\hline
				\textbf{BLUED} & 12 KHz & 1 week & Residential & 1 & 50 & US & {https://tokhub.github.io/dbecd/links/Blued.html} \\
				\hline
				\textbf{UK-DALE} & 16 KHz & 2 years & Residential & 5 & 5 & UK & {https://jack-kelly.com/data/}\\
				\hline
				\textbf{REFIT} & 8 seconds & 2 years & Residential & 20 & 9 & UK & {https://pureportal.strath.ac.uk/en/datasets/refit-electrical-load-measurements}\\
				\hline
				\textbf{AMPds2} & 1 min & 2 years & Residential & 1 & 21 & Canada & {http://dx.doi.org/10.7910/DVN/FIE0S4}\\
				\hline
			\end{tabular}
			\label{tbl:comparison}
		\end{center}
	\end{table*}

	In 2011, the Reference Energy Disaggregation Dataset (REDD) \cite{RN37} became available as the first openly accessible dataset designed specifically to aid NILM research. Following this, the building-Level fully-labeled dataset for Electricity Disaggregation (BLUED) \cite{BLUED} was released in 2012, which contained data from a single household.
	
	The Almanac of Minutely Power dataset (AMPds) \cite{AMPds}, on the other hand, was made public in 2013 and comprised both aggregate and sub-metered power data from a single household. The Almanac of Minutely Power dataset Version 2 (AMPds2) \cite{AMPds2} is another dataset that captures all three primary types of consumption, including electricity, water, and natural gas, over an extended period of 2 years. Furthermore, it provides 11 measurement characteristics for electricity. The data in AMPds2 has been pre-cleaned to ensure consistent and comparable accuracy results among researchers and machine learning algorithms.
	
	The REFIT Electrical Load Measurements \cite{REFIT} dataset is another one that includes cleaned electrical consumption data in Watts for 20 households at both aggregate and appliance levels, timestamped and sampled at 8-second intervals. It is designed to support research into energy conservation and advanced energy services, ranging from non-intrusive appliance load monitoring, demand response measures, tailored energy and retrofit advice, appliance usage analysis, consumption and time-use statistics, and smart home/building automation. Finally, the UK Domestic Appliance-Level Electricity data set (UK-DALE) \cite{UK-DALE} was released, containing data from four households.
	
	The available NILM datasets have varying sample rates ranging from 1 Hz to 100 kHz and cover individual appliances as well as residential complexes. While data collected from individual appliances can be valuable for modeling and training the NILM system, its performance may not be optimal when tested on the entire residential building. Conversely, relying solely on whole-household datasets may not be appropriate for training the algorithms, especially when individual appliance data is unavailable.
	
	Furthermore, certain databases provide primary current and voltage signals, whereas others provide calculated electrical parameters such as active power, reactive power, apparent power, and power factor. However, in order to create an effective NILM system, it is crucial to obtain unprocessed electrical signals in order to extract fundamental and harmonic characteristics.

	\section{Experimental Results}\label{sec:experiments}
	\subsection{Evaluation Metrics}\label{subsec:dataset}
	To evaluate the performance of algorithms in recognizing appliance switching ON or OFF, the classification metrics presented in Eqs. \ref{eq:acc}-\ref{eq:F1} were used. The metrics are calculated using True Positive (TP), True Negative (TN), False Positive (FP), and False Negative (FN). TP represented the number of times a device was correctly recognized as ON, whereas TN represented the number of correctly identified OFF occurrences. FP highlighted instances when ON states were recorded despite the appliance not consuming power. On the other hand, FN displayed the number of OFF occurrences that were incorrectly recognized.
	
	\begin{equation}
		\ \textnormal{Accuracy} = \frac{\textnormal{TP}+\textnormal{TN}}{\textnormal{TP}+\textnormal{TN}+\textnormal{FP}+\textnormal{FN}}
		\label{eq:acc}
	\end{equation}
	
	\begin{equation}
		\ \textnormal{Presicion} = \frac{\textnormal{TP}}{\textnormal{TP}+\textnormal{FP}}
		\label{eq:precision}
	\end{equation}
	
	\begin{equation}
		\ \textnormal{Recall} = \frac{\textnormal{TP}}{\textnormal{TP}+\textnormal{FN}}
		\label{eq:recall}
	\end{equation}
	
	\begin{equation}
		\ F_1 = 2 * \frac{\textnormal{Presicion} * \textnormal{Recall}}{\textnormal{Presicion} + \textnormal{Recall}}
		\label{eq:F1}
	\end{equation}

	The recall metric in Eq. \ref{eq:recall} measures the ratio of correctly identified positive instances (TP) to the total number of positive instances in the dataset. On the other hand, the precision metric in Eq. \ref{eq:precision} represents the ratio of correctly identified positive instances (TP) to the total number of instances identified as positive by the algorithm. The $ F_1 $ score is a weighted mean of precision and recall, which is used to determine the accuracy of the algorithm in identifying appliance states. Higher $ F_1 $ scores indicate better algorithm performance in recognizing appliance state transitions.
	
	The mean absolute error (MAE) and proportion of energy correctly allocated (PECA) metrics in Eqs. \ref{eq:MAE} and \ref{eq:FECA}, respectively, are non-event-based metrics used to evaluate the accuracy of load disaggregation systems in calculating and assigning electricity usage. MAE measures the average absolute difference between the estimated and actual energy usage, while PECA evaluates the percentage of energy correctly allocated to individual appliances.
	
	\begin{equation}
		\ \textnormal{MAE} = 1/T *\sum_{t=1}^{T} | \hat{y}_t^i - {y}_t^i|
		\label{eq:MAE}
	\end{equation}
	
	\begin{equation}
		\ \textnormal{FECA} = 1 - \frac{\sum_{t=1}^{T}\sum_{i=1}^{N}|\hat{y}_t^i - {y}_t^i|}{2\sum_{t=1}^{T}\bar{y}_t} 
		\label{eq:FECA}
	\end{equation}

	In the preceding equations, $\hat{y}_t^i $ and $ {y}_t^i $ are the estimated and ground-truth power of the $ i^{th} $ device at time step t, respectively. Furthermore, $ \bar{y}_t $ is the total power at time t \cite{RN31}.
	\subsection{Performance Evaluation and Comparison Study}\label{subsec:results}
	In this part, the disaggregation findings for the NILM approaches using the REDD, UK-DALE, and REFIT datasets are shown. The performance indicators obtained by executing the tests on these datasets shows that $ F_1 $ produced the best results for refrigerators, air conditioners, freezers, televisions, and washing machines across the three datasets, with values greater than 0.70. Toasters and electronics, on the other hand, have lower $ F_1 $ scores of roughly 0.25 owing to misclassification caused by the non-uniform pattern of these items.
	
	\newcolumntype{L}[1]{>{\raggedright\let\newline\\\arraybackslash\hspace{0pt}}m{#1}}
\newcolumntype{C}[1]{>{\centering\let\newline\\\arraybackslash\hspace{0pt}}m{#1}}
\newcolumntype{R}[1]{>{\raggedleft\let\newline\\\arraybackslash\hspace{0pt}}m{#1}}
\newcommand{\maxnum}{100.00}
\newlength{\maxlen}

\newcommand\tablecellwidthone{1.2cm}

\newcommand{\databarone}[2][blue!25]{%
	\settowidth{\maxlen}{\maxnum}%
	\setlength{\maxlen}{\tablecellwidthone}%
	\FPeval\result{round(#2/\maxnum:4)}%
	\rlap{\color{blue!25}\hspace*{\dimexpr-0.5\tabcolsep+\arrayrulewidth}\rule[-.2\ht\strutbox]{\result\maxlen}{1\ht\strutbox}}%
	\makebox[\dimexpr\maxlen-0.25\tabcolsep+\arrayrulewidth][r]{#2\%}
}

	The accuracy metrics of the findings were compared for published methods: on-line NILM \cite{onlineNILM}, NILM-TK \cite{RN22}, an FHMM implementation; Neural-NILM \cite{RN12}, a DNN adaption for energy estimation. The Neural-NILM used three DNN architectures: i) long short-term memory, ii) de-noising auto-encoders, and iii) rectangles. Rectangle networks, in particular, regress the start-time, end-time, and average power of appliance activation.
	
	In experiments using UK-DALE dataset, a comparison of on-line NILM, NILM-TK and Neural-NILM methods for five appliances (fridge, washing machine, dishwasher, microwave, and kettle) is done. The microwave gives the lowest marks for all three methods. MAE and an $F_1$ score of roughly 195 Watts and 0.01, respectively, are reported by NILM-TK. The best MAE of 6 Watts and an $F_1$ score of 0.21 is shown by the neural-NILM. With an $F_1$ score of about 0.35, the on-line NILM method outperforms the other two. The MAE and $F_1$ scores reported by NILM-TK are roughly 67 watts and 0.55, respectively. The Neural-NILM has an MAE of 18 Watts and an F1 score of 0.82. In terms of energy estimate, the Neural-NILM outperformed the suggested technique, particularly for complicated equipment like dishwashers and washing machines. Nonetheless, the time and computational resources required to train the neural network and generate the models need a large amount of appliance-level data. The on-line NILM technique, on the other hand, may generate appliance models using aggregate data without the necessity for appliance-level sub-metered data.

	\begin{table*}[ht]
		\centering
		\caption{Comparison of Non-Intrusive Load Monitoring (NILM) Methods}
		\label{tab:NILM_Comparison}
		\begin{tabular}{|p{1.5cm}|p{2.5cm}|p{2.5cm}|p{1.5cm}|p{1.5cm}|p{1.5cm}|p{1.5cm}|}
			\hline
			Method & Pros & Cons & Data Requirements & Real-time Performance & Interpretability & Evaluation Performance \\ \hline
			Template Matching & Simple, easy to implement & Requires pre-defined templates for each appliance, limited accuracy & None & Fast & High & Limited by pre-defined templates \\ \hline
			Steady-State Analysis & Accurate for steady-state loads & Limited accuracy for transient loads, requires detailed knowledge of the power system & Detailed power system data & Fast & Medium & Limited by power system model \\ \hline
			HMM and fractional HMM & Can handle transient loads, can identify multiple appliances simultaneously & Requires training data for each appliance, can be computationally expensive & Labeled training data & Slow & Low & Sensitive to initialization and noise \\ \hline
			Deep Neural Networks & High accuracy, can handle complex appliance behavior, can identify multiple appliances simultaneously & Requires a large amount of training data, can be computationally expensive, limited interpretability & Labeled training data & Fast & Low & Dependent on quality and quantity of training data \\ \hline
			Autoencoders & Can handle variable appliance behavior, can identify multiple appliances simultaneously, can learn without labeled data & Requires a large amount of training data, can be computationally expensive & Unlabeled training data & Fast & Low & Dependent on quality and quantity of training data \\ \hline
		\end{tabular}
	\end{table*}

	\section{Conclusion}\label{sec:conclusion}
	The three-phase distribution network often feeds the final residential customers through single phase cables. It is important to make the loads balanced. Demand side management and different optimisation techniques would help on this. However, understanding the contribution of different appliances to energy consumption is beneficial. NILM is demonstrated to be a good approach to this end. The accuracy of NILM depends on the method applied. This paper reviewed some deep learning-based methods which outperform other existing NILM algorithms. The paper compared the results of applying these advanced methods to provide a basis for future implementation. 
	These datasets have public access and are widely used in NILM literature. Several performance criteria are formulated to analyze the performance of the methods.
	\balance
	\bibliographystyle{IEEEtran}
	\bibliography{refs}

\begin{thebibliography}{10}
\providecommand{\url}[1]{#1}
\csname url@samestyle\endcsname
\providecommand{\newblock}{\relax}
\providecommand{\bibinfo}[2]{#2}
\providecommand{\BIBentrySTDinterwordspacing}{\spaceskip=0pt\relax}
\providecommand{\BIBentryALTinterwordstretchfactor}{4}
\providecommand{\BIBentryALTinterwordspacing}{\spaceskip=\fontdimen2\font plus
\BIBentryALTinterwordstretchfactor\fontdimen3\font minus
  \fontdimen4\font\relax}
\providecommand{\BIBforeignlanguage}[2]{{%
\expandafter\ifx\csname l@#1\endcsname\relax
\typeout{** WARNING: IEEEtran.bst: No hyphenation pattern has been}%
\typeout{** loaded for the language `#1'. Using the pattern for}%
\typeout{** the default language instead.}%
\else
\language=\csname l@#1\endcsname
\fi
#2}}
\providecommand{\BIBdecl}{\relax}
\BIBdecl

\bibitem{RN1}
A.~Garulli, S.~Paoletti, and A.~Vicino, ``Models and techniques for electric
  load forecasting in the presence of demand response,'' \emph{IEEE
  Transactions on Control Systems Technology}, vol.~23, no.~3, pp. 1087--1097,
  2014.

\bibitem{RN2}
A.~Estebsari and R.~Rajabi, ``Single residential load forecasting using deep
  learning and image encoding techniques,'' \emph{Electronics}, vol.~9, no.~1,
  p.~68, 2020.

\bibitem{RN3}
R.~Rajabi and A.~Estebsari, ``Deep learning based forecasting of individual
  residential loads using recurrence plots,'' in \emph{2019 IEEE Milan
  PowerTech}.\hskip 1em plus 0.5em minus 0.4em\relax IEEE, 2019, pp. 1--5.

\bibitem{EEEIC_PriceGRU}
N.~Rezaei, R.~Rajabi, and A.~Estebsari, ``Electricity price forecasting model
  based on gated recurrent units,'' in \emph{2022 IEEE International Conference
  on Environment and Electrical Engineering and 2022 IEEE Industrial and
  Commercial Power Systems Europe (EEEIC/I\&CPS Europe)}, 2022, pp. 1--5.

\bibitem{RN4}
T.~Wang, B.~Shen, C.~H. Springer, and J.~Hou, ``What prevents us from taking
  low-carbon actions? a comprehensive review of influencing factors affecting
  low-carbon behaviors,'' \emph{Energy Research \& Social Science}, vol.~71, p.
  101844, 2021.

\bibitem{RN5}
I.~Ishak, N.~S. Othman, and N.~H. Harun, ``Forecasting electricity consumption
  of malaysia’s residential sector: Evidence from an exponential smoothing
  model,'' \emph{F1000Research}, vol.~11, no.~54, p.~54, 2022.

\bibitem{RN12}
D.~Shi, R.~Li, R.~Shi, and F.~Li, ``Analysis of the relationship between load
  profile and weather condition,'' in \emph{2014 IEEE PES General Meeting|
  Conference \& Exposition}.\hskip 1em plus 0.5em minus 0.4em\relax IEEE, 2014,
  pp. 1--5.

\bibitem{RN13}
M.~U. Fahad and N.~Arbab, ``Factor affecting short term load forecasting,''
  \emph{Journal of Clean Energy Technologies}, vol.~2, no.~4, pp. 305--309,
  2014.

\bibitem{RN18}
R.~Bonfigli, S.~Squartini, M.~Fagiani, and F.~Piazza, ``Unsupervised algorithms
  for non-intrusive load monitoring: An up-to-date overview,'' in \emph{2015
  IEEE 15th international conference on environment and electrical engineering
  (EEEIC)}.\hskip 1em plus 0.5em minus 0.4em\relax IEEE, 2015, pp. 1175--1180.

\bibitem{RN19}
O.~Parson, S.~Ghosh, M.~Weal, and A.~Rogers, ``Non-intrusive load monitoring
  using prior models of general appliance types,'' in \emph{Twenty-Sixth AAAI
  Conference on Artificial Intelligence}, 2012.

\bibitem{RN20}
------, ``An unsupervised training method for non-intrusive appliance load
  monitoring,'' \emph{Artificial Intelligence}, vol. 217, pp. 1--19, 2014.

\bibitem{RN21}
J.~Z. Kolter and T.~Jaakkola, ``Approximate inference in additive factorial
  hmms with application to energy disaggregation,'' in \emph{Artificial
  intelligence and statistics}.\hskip 1em plus 0.5em minus 0.4em\relax PMLR,
  2012, pp. 1472--1482.

\bibitem{RN14}
X.~Sun, P.~B. Luh, K.~W. Cheung, W.~Guan, L.~D. Michel, S.~Venkata, and M.~T.
  Miller, ``An efficient approach to short-term load forecasting at the
  distribution level,'' \emph{IEEE Transactions on Power Systems}, vol.~31,
  no.~4, pp. 2526--2537, 2015.

\bibitem{RN22}
H.~Gon{\c{c}}alves, A.~Ocneanu, M.~Berg{\'e}s, and R.~Fan, ``Unsupervised
  disaggregation of appliances using aggregated consumption data,'' in
  \emph{The 1st KDD workshop on data mining applications in sustainability
  (SustKDD)}, 2011.

\bibitem{RN15}
J.~Zheng, C.~Xu, Z.~Zhang, and X.~Li, ``Electric load forecasting in smart
  grids using long-short-term-memory based recurrent neural network,'' in
  \emph{2017 51st Annual Conference on Information Sciences and Systems
  (CISS)}.\hskip 1em plus 0.5em minus 0.4em\relax IEEE, 2017, pp. 1--6.

\bibitem{RN23}
M.~Zhuang, M.~Shahidehpour, and Z.~Li, ``An overview of non-intrusive load
  monitoring: Approaches, business applications, and challenges,'' in
  \emph{2018 international conference on power system technology
  (POWERCON)}.\hskip 1em plus 0.5em minus 0.4em\relax IEEE, 2018, pp.
  4291--4299.

\bibitem{RN24}
L.~Mauch and B.~Yang, ``A new approach for supervised power disaggregation by
  using a deep recurrent lstm network,'' in \emph{2015 IEEE Global Conference
  on Signal and Information Processing (GlobalSIP)}.\hskip 1em plus 0.5em minus
  0.4em\relax IEEE, 2015, pp. 63--67.

\bibitem{RN25}
J.~Kelly and W.~Knottenbelt, ``Neural nilm: Deep neural networks applied to
  energy disaggregation,'' in \emph{Proceedings of the 2nd ACM international
  conference on embedded systems for energy-efficient built environments},
  2015, pp. 55--64.

\bibitem{RN26}
C.~Shin, S.~Joo, J.~Yim, H.~Lee, T.~Moon, and W.~Rhee, ``Subtask gated networks
  for non-intrusive load monitoring,'' in \emph{Proceedings of the AAAI
  Conference on Artificial Intelligence}, vol.~33, no.~01, 2019, pp.
  1150--1157.

\bibitem{RN27}
C.~Zhang, M.~Zhong, Z.~Wang, N.~Goddard, and C.~Sutton, ``Sequence-to-point
  learning with neural networks for non-intrusive load monitoring,'' in
  \emph{Proceedings of the AAAI Conference on Artificial Intelligence},
  vol.~32, no.~1, 2018.

\bibitem{RN10}
E.~Busseti, I.~Osband, and S.~Wong, ``Deep learning for time series modeling,''
  \emph{Technical report, Stanford University}, pp. 1--5, 2012.

\bibitem{RN29}
S.~Makonin, F.~Popowich, I.~V. Baji{\'c}, B.~Gill, and L.~Bartram, ``Exploiting
  hmm sparsity to perform online real-time nonintrusive load monitoring,''
  \emph{IEEE Transactions on smart grid}, vol.~7, no.~6, pp. 2575--2585, 2015.

\bibitem{RN30}
H.~Kim, M.~Marwah, M.~Arlitt, G.~Lyon, and J.~Han, ``Unsupervised
  disaggregation of low frequency power measurements,'' in \emph{Proceedings of
  the 2011 SIAM international conference on data mining}.\hskip 1em plus 0.5em
  minus 0.4em\relax SIAM, 2011, pp. 747--758.

\bibitem{RN31}
J.~Kelly and W.~Knottenbelt, ``Neural nilm: Deep neural networks applied to
  energy disaggregation,'' in \emph{Proceedings of the 2nd ACM international
  conference on embedded systems for energy-efficient built environments},
  2015, pp. 55--64.

\bibitem{RN32}
P.~P.~M. do~Nascimento, ``Applications of deep learning techniques on nilm,''
  \emph{Diss. Universidade Federal do Rio de Janeiro}, 2016.

\bibitem{RN33}
S.~Henriet, U.~{\c{S}}im{\c{s}}ekli, B.~Fuentes, and G.~Richard, ``A generative
  model for non-intrusive load monitoring in commercial buildings,''
  \emph{Energy and Buildings}, vol. 177, pp. 268--278, 2018.

\bibitem{onlineNILM}
M.~A. Mengistu, A.~A. Girmay, C.~Camarda, A.~Acquaviva, and E.~Patti, ``A
  cloud-based on-line disaggregation algorithm for home appliance loads,''
  \emph{IEEE Transactions on Smart Grid}, vol.~10, no.~3, pp. 3430--3439, 2019.

\bibitem{MDPI_CNN}
L.~Massidda, M.~Marrocu, and S.~Manca, ``Non-intrusive load disaggregation by
  convolutional neural network and multilabel classification,'' \emph{Applied
  Sciences}, vol.~10, no.~4, 2020.

\bibitem{RN34}
B.~Zhao, L.~Stankovic, and V.~Stankovic, ``On a training-less solution for
  non-intrusive appliance load monitoring using graph signal processing,''
  \emph{IEEE Access}, vol.~4, pp. 1784--1799, 2016.

\bibitem{RN35}
T.~Kriechbaumer, D.~Jorde, and H.-A. Jacobsen, ``Waveform signal entropy and
  compression study of whole-building energy datasets,'' in \emph{Proceedings
  of the Tenth ACM International Conference on Future Energy Systems}, 2019,
  pp. 58--67.

\bibitem{RN36}
L.~Wen, K.~Zhou, and S.~Yang, ``Load demand forecasting of residential
  buildings using a deep learning model,'' \emph{Electric Power Systems
  Research}, vol. 179, p. 106073, 2020.

\bibitem{RN37}
J.~Z. Kolter and M.~J. Johnson, ``Redd: A public data set for energy
  disaggregation research,'' in \emph{Workshop on data mining applications in
  sustainability (SIGKDD), San Diego, CA}, vol.~25, no. Citeseer, 2011, pp.
  59--62.

\bibitem{BLUED}
K.~Anderson, A.~Ocneanu, D.~Benitez, D.~Carlson, A.~Rowe, and M.~Berges,
  ``Blued: A fully labeled public dataset for event-based non-intrusive load
  monitoring research,'' in \emph{Proceedings of the 2nd KDD workshop on data
  mining applications in sustainability (SustKDD)}, vol.~7.\hskip 1em plus
  0.5em minus 0.4em\relax ACM New York, 2012, pp. 1--5.

\bibitem{AMPds}
S.~Makonin, F.~Popowich, L.~Bartram, B.~Gill, and I.~V. Bajić, ``Ampds: A
  public dataset for load disaggregation and eco-feedback research,'' in
  \emph{2013 IEEE Electrical Power \& Energy Conference}, 2013, pp. 1--6.

\bibitem{AMPds2}
\BIBentryALTinterwordspacing
S.~Makonin, ``Ampds2: The almanac of minutely power dataset (version 2),''
  2021. [Online]. Available: \url{https://dx.doi.org/10.21227/c2rq-bx53}
\BIBentrySTDinterwordspacing

\bibitem{REFIT}
D.~Murray, L.~Stankovic, and V.~Stankovic, ``An electrical load measurements
  dataset of united kingdom households from a two-year longitudinal study,''
  \emph{Scientific data}, vol.~4, no.~1, pp. 1--12, 2017.

\bibitem{UK-DALE}
J.~Kelly and W.~Knottenbelt, ``The uk-dale dataset, domestic appliance-level
  electricity demand and whole-house demand from five uk homes,''
  \emph{Scientific data}, vol.~2, no.~1, pp. 1--14, 2015.

\end{thebibliography}
\end{document}